\documentclass{emulateapj}
\usepackage{amssymb,amsfonts,amsmath,amstext,amsgen,amsopn,amsxtra,indentfirst,lscape}
\shorttitle{Extinction Law in Intermediate-$z$ Galaxy}
\shortauthors{Heng et al.}
\begin{document}

\title{A Direct Measurement of the Dust Extinction Curve in an
Intermediate-Redshift Galaxy}

\author{Kevin Heng\altaffilmark{1}, Davide Lazzati\altaffilmark{2},
Rosalba Perna\altaffilmark{2}, Peter Garnavich\altaffilmark{3}, Alberto Noriega-Crespo\altaffilmark{4}, David Bersier\altaffilmark{5}, Thomas Matheson\altaffilmark{6} \& Michael Pahre\altaffilmark{7}}

\altaffiltext{1}{Institute for Advanced Study, School of Natural
Sciences, Einstein Drive, Princeton, NJ 08540; heng@ias.edu}

\altaffiltext{2}{JILA, University of Colorado, 440 UCB, 
Boulder, CO 80309-0440}

\altaffiltext{3}{University of Notre Dame, Department of Physics, 
225 Nieuwland Science Hall, Notre Dame, IN 46556}

\altaffiltext{4}{Spitzer Science Center, California Institute of 
Technology, Pasadena, CA 91125}

\altaffiltext{5}{Astrophysics Research Institute, Liverpool John 
Moores University, Twelve Quays House, Egerton Wharf, Birkenhead, 
CH41 1LD, United Kingdom}

\altaffiltext{6}{NOAO, 950 North Cherry Avenue, Tucson, AZ 85719}

\altaffiltext{7}{Harvard Smithsonian Center for Astrophysics, 60 Garden 
Street, Cambridge, MA 02138}

\begin{abstract}
We present a proof-of-concept study that dust extinction curves can be extracted from the infrared (IR), optical, ultraviolet (UV) and X-ray afterglow observations of GRBs without assuming known extinction laws.  We focus on GRB 050525A ($z = 0.606$), for which we also present IR observations from the {\it Spitzer Space Telescope} at $t = t_{\rm IR} \approx 2.3$ days post-burst.  We construct the spectral energy distribution (SED) of the afterglow at $t = t_{\rm IR}$ and use it to derive the dust extinction curve of the host galaxy in 7 optical/UV wavebands.  By comparing our derived extinction curve to known templates, we see that the Galactic or Milky Way extinction laws are disfavored versus the Small and Large Magellanic Cloud (SMC and LMC) ones, but that we cannot rule out the presence of a LMC-like 2175 \AA\ bump in our extinction curve.  The dust-to-gas ratio present within the host galaxy of GRB 050525A is similar to that found in the LMC, while about 10 to 40\% more dust is required if the SMC template is assumed.  Our method is useful to observatories that are capable of simultaneously observing GRB afterglows in multiple wavebands from the IR to the X-ray.
\end{abstract}


\section{Introduction}
\label{sect:intro}
Extinction\footnote{In this paper, we refer to ``extinction'' as the absorption and scattering of photons from a single line of sight.  ``Attenuation'' refers to the aggregate effect of possibly complex source and dust distributions, including multiple scattering and  the scattering of photons back into the line of sight.} of light plays a major role in the observation of any astronomical object, but even more so for intermediate- and high-redshift ones due to the intrinsically fainter signals and our poorer understanding of the dust physics involved.  In the optical range of wavelengths, dust grains are the dominant contributor to extinction.  High-quality extinction curves have been constructed for our Galaxy (the Milky Way; MW), as well as for the Small and Large Magellanic Clouds (SMC and LMC; e.g., Cardelli, Clayton \& Mathis 1989; Pei 1992; Weingartner \& Draine 2001).  

Despite the importance of the subject, surprisingly little is presently known about the extinction properties of dust grains in the universe at redshifts of $z>0$.  Some constraints on the properties of dust have been obtained for active galactic nuclei (AGN; Laor \& Draine 1993) and starburst galaxies at low redshifts (Calzetti 1997).  Nevertheless, our knowledge of the composition and grain size distribution of dust in intermediate- and high-$z$ galaxies is very limited.  The only constraints set on some intermediate-$z$ galaxies ($z\sim 0.5$ to 0.9) were obtained through the analysis of multi-imaged quasars (QSOs) behind a lensing galaxy (Falco et al. 1999; Motta et al. 2002; Mu\~{n}oz et al. 2004). These studies utilize the fact that the lensed images have the same intrinsic spectrum.  The extinction curves, which are measured and characterized by a Cardelli, Clayton \& Mathis (1989) profile, range from flat (``gray'') to steep, SMC-like profiles, both with and without the 2175\AA\ feature.  

Other studies have derived attenuation curves, obtained by substituting a source population --- for which an average, intrinsic spectrum can be safely assumed --- for the point-like source.  Notable results were achieved by Calzetti (1997) for starburst galaxies and by Johnson et al. (2007) for a sample of 1000 {\it Sloan Digital Sky Survey} (SDSS) galaxies.  Strikingly, the former and latter studies yielded gray and SMC-like attenuation (or ``effective extinction'') curves, respectively.  These studies leave open the question of whether the inferred diversity stems from intrinsic differences in the dust properties or is an effect of hidden (and incorrect) assumptions in the utilized methods.

Our relatively poor knowledge of dust physics at $z>0$ makes it difficult to derive the fundamental properties of intermediate- and high-$z$ galaxies such as the star formation rate (or at least its obscured fraction), the metallicity of the interstellar medium
(ISM), the rate of supernova (SN) explosions, and the extinction-corrected luminosity of any galactic source of light.  Our level of understanding of dust physics in our Galaxy shows us what can be learned if high-quality extinction curves
are available.  Studying the extinction properties of the ISM in the MW has provided insight into dust composition, which has in turn taught us about grain coagulation and destruction, metal depletion (into dust), and grain polarization properties  (e.g., Mathis, Rumpl \& Nordsieck 1977; Weingartner \& Draine 2001; Draine \& Li 2007).  These pioneering studies have set the stage for asking whether MW dust physics is unique to our Galaxy or evolves with redshift.  Differences between the MW, SMC and LMC extinction laws suggest that grain properties probably vary with the local radiation field, metallicity and galactic mass, but the lack of more examples has hampered any further conclusions.

Measuring the extinction curve at intermediate- or high-$z$ is a formidable task, due to the rarity of sources with known spectra and luminosities.  The discovery of $\gamma$-ray bursts (GRBs) and their afterglows, about a decade ago, has opened a new window into the study of dust physics.  In particular, long-duration GRBs are believed to occur in dusty environments, since it is thought that they result from the deaths of very massive stars.  Their afterglows have featureless, power-law spectra (e.g., Liang et al. 2007) and intense luminosities ($\sim 10^{48}$ to $10^{50}$ erg s$^{-1}$), easily outshining a typical QSO ($\sim 10^{46}$ erg s$^{-1}$).  It has been previously demonstrated that the study of the infrared (IR), optical, ultraviolet (UV) and X-ray afterglows of a GRB can place constraints on the extinction by dust present within the host galaxy (e.g., Price et al. 2002; Starling et al. 2007).  However, these studies lacked an unbiased measurement of the unabsorbed spectrum and its normalization, restricting the analyses to only fitting known extinction laws to the data, rather than directly measure the dust extinction.

In this paper, we pioneer a method that is similar in spirit to the previous studies, but uses less assumptions.  For our study, we focus on GRB 050525A, the first detection of a GRB at mid-IR wavelengths with the {\it Spitzer Space Telescope} (Garnavich et al. 2005).  We demonstrate that by constructing a spectral energy distribution (SED) of the GRB afterglow which includes IR and X-ray data points, one can derive an extinction curve without having to assume a known dust extinction law.  In other words, our method allows a template-independent calibration of the extinction curve.  We emphasize that such an approach is well-suited for observatories that are capable of simultaneously observing GRB afterglows in multiple wavebands (e.g., {\it GROND}).

In \S\ref{sect:observations}, we describe the IR, optical, UV and X-ray observations used in our study.  Our {\it Spitzer} IR observations at $t = t_{\rm IR} \approx 2.3$ days are not contemporaneous with the optical, UV and X-ray observations used, which motivates a description, in \S\ref{sect:lightcurves}, of our light curve fitting procedure that interpolates for the optical, UV and X-ray SED data points at $t = t_{\rm IR}$.  Using our constructed SED, we derive an extinction curve for the host galaxy of GRB 050525A in \S\ref{sect:results}; we further analyze its dust-to-gas ratio if known extinction laws are fitted to our curve.  Finally, we compare our results to that of previous studies, discuss their implications and present opportunities for future work in \S\ref{sect:discussion}.

\section{Observations}
\label{sect:observations}

\subsection{Optical, Ultraviolet \& X-ray Data from Literature}

GRB 050525A occurred at 00:02:53 UT on 2005 May 25th (Band et al. 2005) at a redshift of $z=0.606$ (Foley et al. 2005).  Optical and UV observations of its afterglow, in the $UVW2$ (198 nm), $UVM2$ (220 nm), $UVW1$ (260 nm), $U$, $B$
and $V$ bands were taken by Blustin et al. (2006, hereafter B06) using the {\it Swift UVOT}.  B06 also performed {\it Swift XRT}
measurements in the 2 to 10 keV bands.  The optical and UV observations were corrected for Galactic extinction using the extinction data of Schlegel, Finkbeiner \& Davis (1998) and Pei (1992).  For corrections to the X-ray observations, the Galactic hydrogen column density was taken to be $N^{(g)}_{\rm H} = 9.0 \times 10^{20}$ cm$^{-2}$; the hydrogen-equivalent column density at $z=0.606$ was estimated to be about 2---$3 \times 10^{21}$ cm$^{-2}$.  (See \S2.3 of B06.)

The $R$ band data is taken predominantly
from Klotz et al. (2005) and Della Valle et al. (2006, hereafter D06) for early ($\lesssim 100$ min) and
late ($\gtrsim 100$ min) times, respectively.  These are supplemented
by the observations of Cobb \& Bailyn (2005), Homewood et al. (2005)
and Haislip et al. (2005).  GRB 050525A is thought to be observed on-axis and is associated with
supernova (SN) 2005nc; D06 proposed that a bump in the $R$ band light curve at
$\sim$ 20 days is due to a SN contribution peaking at between 10 and
20 days.  We therefore consider $R$ band data only up to about 5 days.  The $R$ band magnitudes are converted to flux densities taking
into account the Galactic extinction of $A^{(g)}_R = 0.25$ mag towards
GRB 050525A (Schlegel, Finkbeiner \& Davis 1998).

To account for systematic errors inherent in using data from different observatories, we enforce a minimum error of 5\% ($\sim0.05$ mag) for the optical/UV data.

\subsection{Infrared Observations from {\it Spitzer}}

Our IR observations were triggered using a ``target-of-opportunity''
(ToO) program on the {\it Spitzer Space Telescope} (Garnavich et
al. 2005).  Imaging and spectroscopy began on May 27.264 (UT) at
$t=t_{\rm IR} = 1.95 \times 10^5$s $\approx 2.3$ days post-burst.  We obtained measurements (Table \ref{tab:spitzer}) from the 3.6, 4.5, 5.8 and 8.0 $\mu$m {\it IRAC} channels, and
from the 24.0 $\mu$m {\it MIPS} channel, using the latest {\it Spitzer Science Center} pipeline products (S16; 2007 Dec).  The most current flux calibration was used (Engelbracht et al. 2007).  The data required very little post-processing beyond the standard products, because the source was placed at the center of the array (Engelbracht et al. 2007).  An aperture photometry measurement was carried out using a 7\arcsec radius at the position of the source (RA = 18h32m32.58s, DEC = 26d20m22.5s, J2000) that required an aperture correction of 2.05 (Table 1 in Engelbracht et al. 2007) to recover all of the flux.  The selected aperture radius reduces the light contamination by a nearby source; for the same reason, the background was measured on a comparable area $\sim 10\arcsec$ away from GRB 050525A.

\subsection{Host Galaxy Contribution \& Extinction}
\label{subsect:host}

The optical and ultraviolet contributions from the host galaxy are at the $\lesssim 1\%$ level on average, especially at early times (B06).  There is a definitive measurement of the host galaxy contribution only in the $R$ band: $R=25.2$ mag (D06).  Since the X-ray and other optical/UV contributions are expected to be small, we ignore them in our analysis.  However, it is conceivable that the host galaxy and GRB afterglow emissions might be of comparable strengths in the 24.0 $\mu$m band (e.g., Zheng et al. 2006; Martin et al. 2007).  Unfortunately, the host galaxy contribution at 24.0 $\mu$m cannot be empirically determined with just a single epoch of IR observations.  We therefore consider two bracketing scenarios: Models A and B.  In Model A, the measured 24.0 $\mu$m flux is assumed to be entirely from the GRB afterglow.  In Model B, we reduce the 24.0 $\mu$m flux measurement by a factor of 2 when constructing a spectral energy distribution (SED) at $t=t_{\rm IR} \approx 2.3$ days post-burst (see \S\ref{sect:results}).  

We further assume that dust extinction by the host galaxy in the IR and X-ray is negligible.  We show later (in \S\ref{subsect:dustgas}) that the derived visual extinction is $A_V \sim 0.2$, which implies, for example, that dust extinction in the 3.6 $\mu$m band is negligibly small, since it is $\sim 5\%$ of that in the $V$ band (B. Schmidt 2008, private communication).  However, there may be a gray component to the attenuation (e.g., Perley et al. 2008), which we do not consider.

\section{Light Curve Fitting}
\label{sect:lightcurves}

Due to the lack of optical and UV data taken at the same time as the IR
observations, we evaluate the fluxes at $t=t_{\rm IR} \approx 2.3$ days by
modeling the light curves according to the standard afterglow model (Sari, Piran \& Narayan 1998; Panaitescu \& Kumar 2000).  The thorough sampling of the $R$ band light curve at late times allows for an interpolation, and not a much more dangerous extrapolation, of the data.  We assume that the optical and UV radiation is powered by the same population of electrons, thus allowing for the temporal evolution of the optical and UV light curves to be the same.

We fit to the data a double power law of the form
\begin{equation}
F_\nu\left(\nu,t\right) = F_\nu\left( \nu,t_{\rm{IR}} \right) ~\left[ \frac{
\left( t_{\rm{IR}}/t_b \right)^{s \alpha_1} + \left( t_{\rm{IR}}/t_b
\right)^{s \alpha_2} }{ \left( t/t_b \right)^{s \alpha_1} + \left(
t/t_b \right)^{s \alpha_2} } \right]^{1/s},
\end{equation}
where $\alpha_1$ and $\alpha_2$ are the asymptotic power law indices
at $t \ll t_b$ and $t \gg t_b$, respectively, and $\nu$ is the observed frequency.  The index $s$
controls the sharpness of the transition at $t=t_b$; we adopt $s=3$ throughout, since it turns out to be
irrelevant for our results (see Liang et al. 2007).  The transition at $t=t_b$ may be interpreted as a jet break; its existence at $\sim 0.16$ to 0.2 days has been discussed in the literature (B06; Liang et
al. 2007), albeit with some controversy (Sato et al. 2007).  D06
reported a jet break at about 0.3 days from an analysis of the $R$
band data only.

\begin{figure}
\begin{center}
\includegraphics[width=5in]{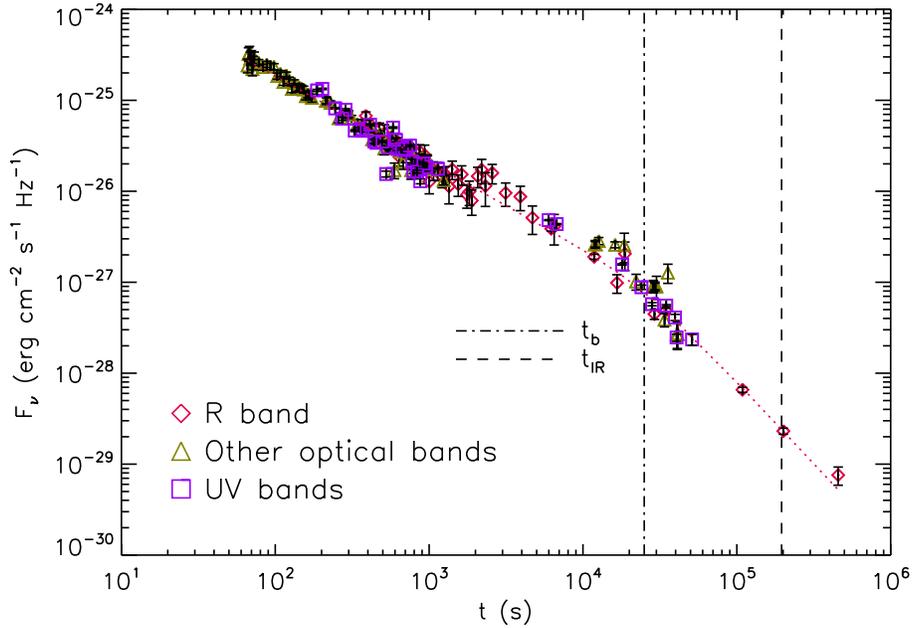}
\end{center}
\caption{Combined optical and ultraviolet light curves, where data points from the other 6 bands are scaled to the $R$ band model fit (dotted curve).  Note that the scaling is used only for visualization and was not employed in the fitting algorithm.}
\label{fig:combined}
\end{figure}

For the optical and UV data, we fit for
$F_\nu (\nu,t_{\rm IR})$, $\alpha_1$, $\alpha_2$ and $t_b$ in the 7 wavebands
simultaneously (i.e., 10 parameters; Fig. \ref{fig:combined}).  The optical/UV fit is largely
constrained by the $R$ band, for which we have the most data points
(44).  The reduced chi-square value for the combined optical/UV fit is
$\chi^2_r = 310/142 = 2.2$. Since the excess $\chi^2$ is evenly
distributed at all times and frequencies, we believe this somewhat
large value of $\chi^2_r$ to be due to either an
under-estimation of the data errors or to intrinsic high-frequency
fluctuations in the light curve. This will be appropriately taken into
account when computing the uncertainties on the fitted parameters.

We perform an analogous fit to the X-ray data, which yields $\chi^2_r =
68.5/54 = 1.3$.  Upon obtaining a model fit to the X-ray flux, we
convert it into a flux density, $F^{({\rm x})}_\nu$.  Following B06,
we assume a power law with a spectral index $\beta_{\rm x} \approx
-0.9$, $F^{({\rm x})}_\nu \propto \nu^{\beta_{\rm x}}$.  

The evaluation of the uncertainties in the optical and UV fluxes at
$t=t_{\rm IR}$ is of particular importance for our study. For each
parameter $P$, we compute the marginalized chi-square, $\chi^2_m (P)$.
The 1-$\sigma$ errors, $P_\sigma$, are determined from the condition:
\begin{equation}
\chi^2_m \left( P_\sigma \right) - {\rm min} 
\{ \chi^2_m \left( P \right) \} = \chi^2_r.
\end{equation}
The preceding condition is stronger than the usual one (where the difference between
the LHS quantities is set to unity), because it takes into account the
uncertainties due to random fluctuations in the data that cause the minimum $\chi^2$ to have a higher than optimal value. The resulting fluxes and their uncertainties at
$t = t_{\rm IR}$ are reported in Table \ref{tab:spitzer}.

For the optical and UV data, we obtain: $\alpha_{1,{\rm ov}} = 0.95
\pm 0.01$, $\alpha_{2,{\rm ov}} = 1.80^{+0.05}_{-0.06}$ and $t_{b,{\rm
ov}} = (2.50^{+0.20}_{-0.31}) \times 10^4$~s $\approx 0.3$ days; our
$t_{b,{\rm ov}}$ value is consistent with the finding of D06.  For the
X-ray data, we get: $\alpha_{1,{\rm x}} = 0.90^{+0.02}_{-0.04}$,
$\alpha_{2,{\rm x}} = 1.56 \pm 0.04$ and $t_{b,{\rm x}} =
(1.46^{+0.25}_{-0.31}) \times 10^3$ s $\approx 0.4$ hours.

\section{Results}
\label{sect:results}

\subsection{Constructing the SED at 2.3 Days Post-Burst}
\label{subsect:sed}

\begin{figure}
\begin{center}
\includegraphics[width=5in]{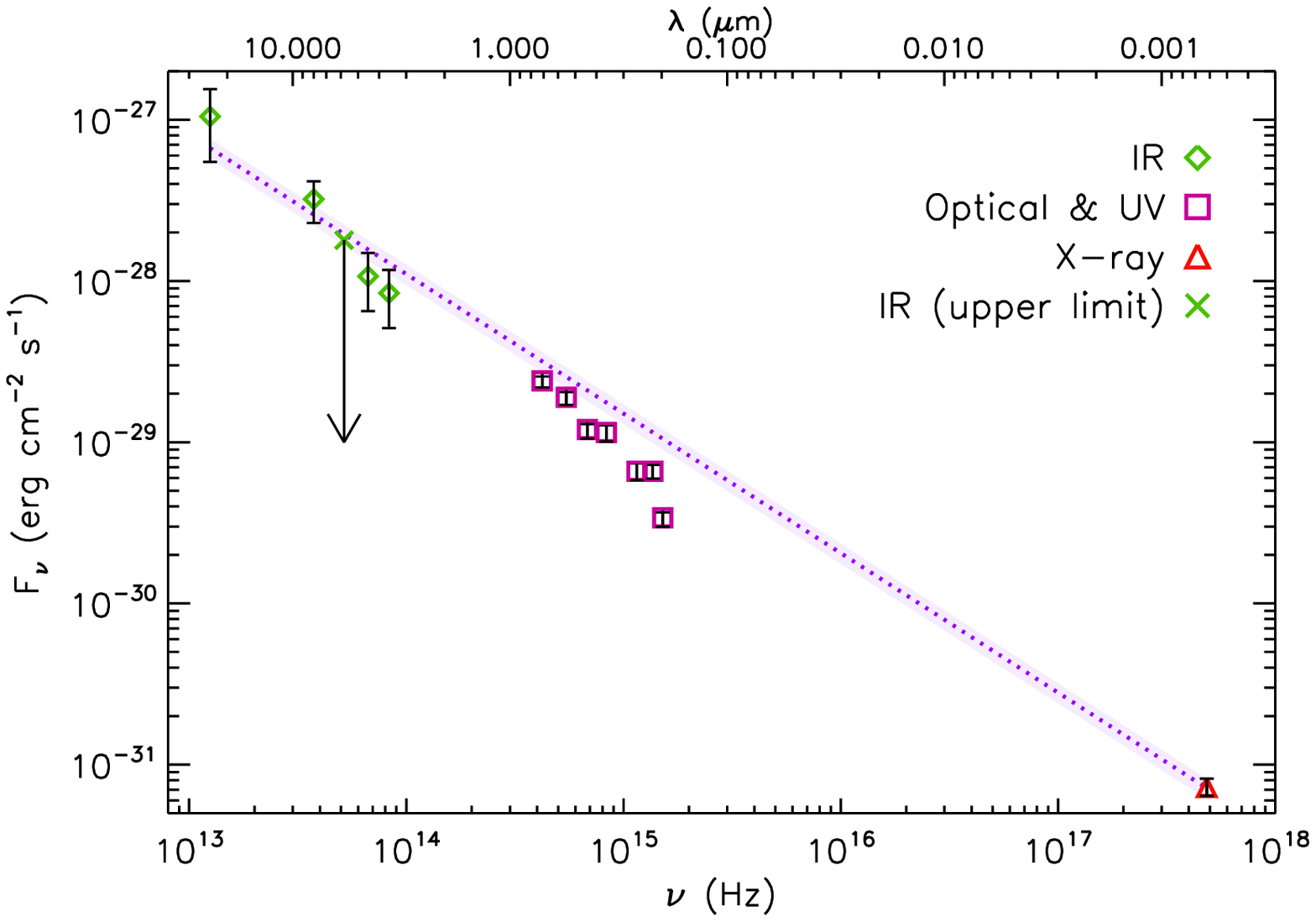}
\includegraphics[width=5in]{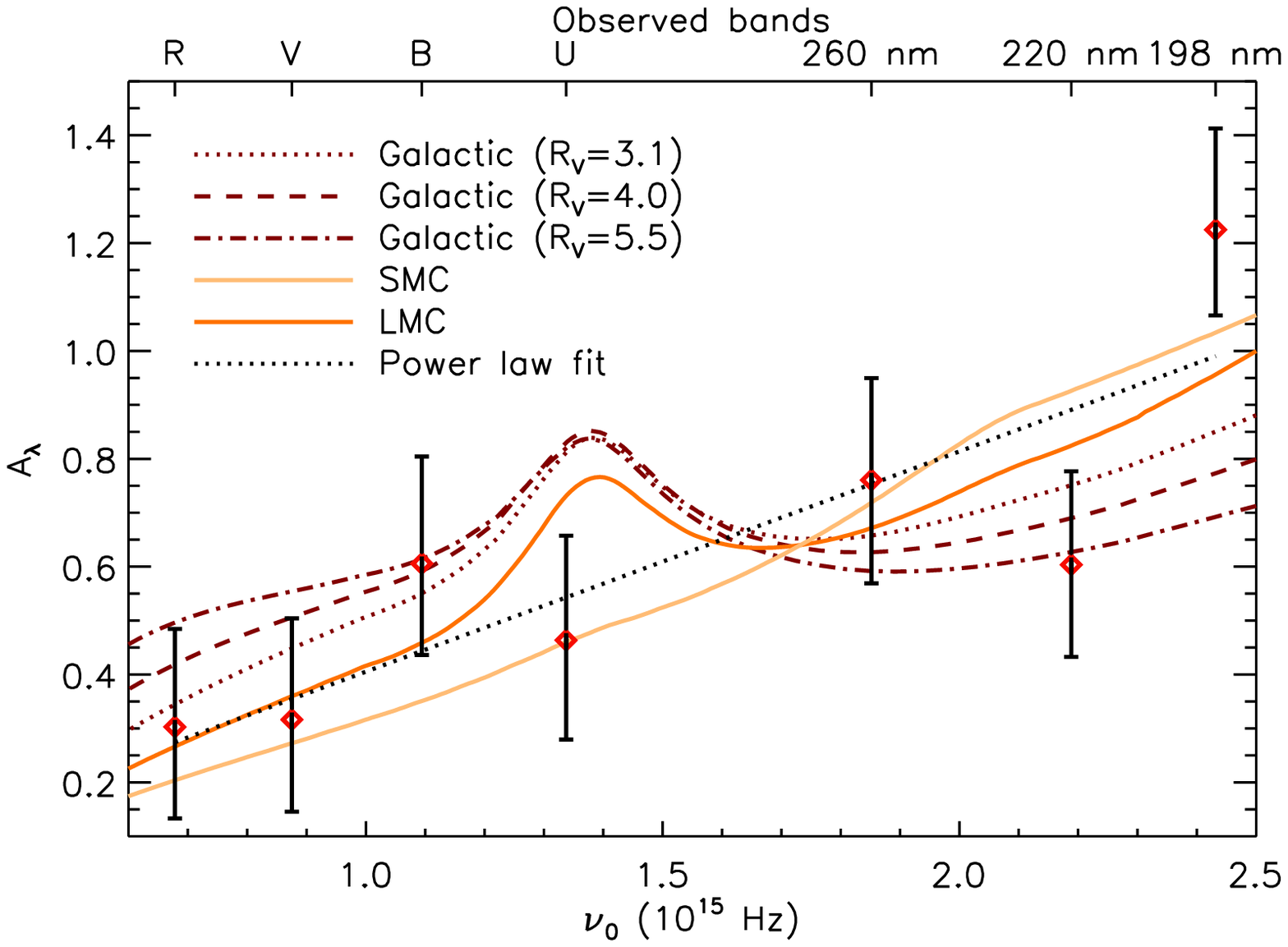}
\end{center}
\caption{Derived spectral energy distribution (SED) and extinction curve for Model A (see text).  Top: SED of the afterglow of GRB 050525A at $t=t_{\rm IR} \approx 2.3$ days after the burst as a function of the observed frequency, $\nu$.  A power law with an index of $\beta = -0.87 \pm 0.02$ is fitted to the infrared and X-ray data points.  Bottom: Extinction curve, $A_\lambda$, for the host galaxy of GRB 050525A as a function of the rest frame frequency, $\nu_0$.  Shown are fits to known dust extinction templates: the Small and Large Magellanic Clouds (SMC and LMC), the Milky Way (MW) for different $R_V$ values and a power law fit (with an index of $\delta = 1.01 \pm 0.31$).}
\label{fig:model_a}
\end{figure}

\begin{figure}
\begin{center}
\includegraphics[width=5in]{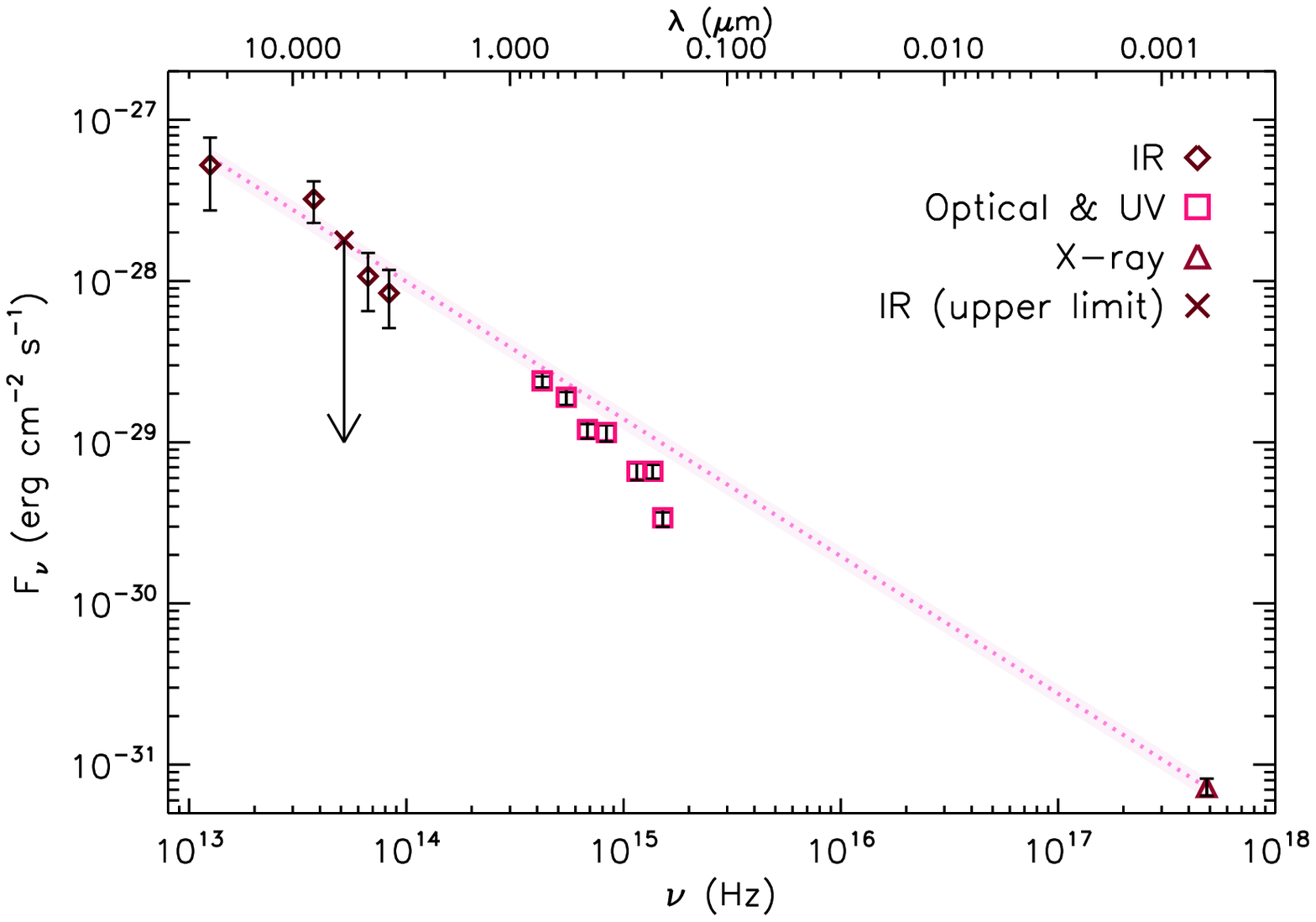}
\includegraphics[width=5in]{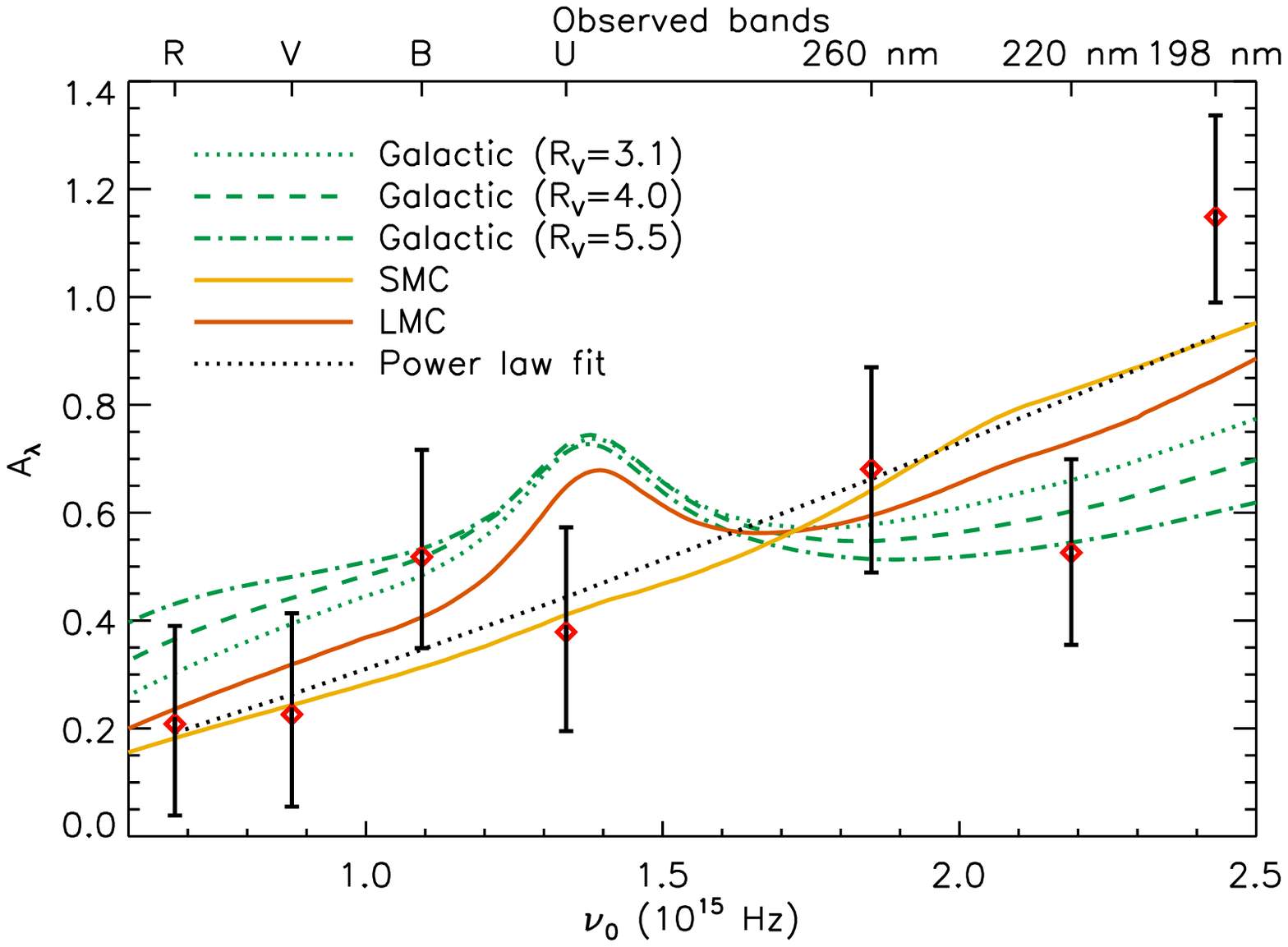}
\end{center}
\caption{Same as Fig. \ref{fig:model_a}, but for Model B (see text).}
\label{fig:model_b}
\end{figure}

Using the optical, UV and X-ray fits, as well as the IR data and upper
limit, we construct the SED for the afterglow at $t=t_{\rm IR}$ for both Models A and B (Figs. \ref{fig:model_a} and \ref{fig:model_b}).    In the standard fireball model of Sari, Piran \& Narayan (1998), the SED at $t=t_{\rm IR}$ transitions from a power law with an index of $(1-p)/2$ to one with an index of $-p/2$ at the cooling frequency of
\begin{equation}
\nu_c = 3.8 \times 10^{16} \mbox{ Hz}
~\epsilon^{-3/2}_{\rm{B},-3} ~\left(\frac{g_c}{0.112}\right) ~n^{-1}_0
~E^{-1/2}_{{\rm iso},53},
\end{equation}
where $\epsilon_{\rm{B},-3}$ is the fraction of the shock energy
injected into the magnetic field in units of $10^{-3}$, $n_0$ is the
density of the circum-burst material in units of 1 cm$^{-3}$, $E_{\rm
iso} = E_{{\rm iso},53} 10^{53}$ erg is the isotropic equivalent energy, and $g_c = (p -
0.46)\exp{(-1.16p)}$ is a function of the power law index $p \approx
2.5$ of the electron distribution.  (See also Shao \& Dai 2005.)  With
our SED, we do not see evidence for a cooling break, consistent with
the findings of B06, who noted the absence of a break for the SED at
$t=25,000$ s.

The observed absence of a synchrotron cooling break in the SED allows us to fit a single power law to it, using only the IR and X-ray flux densities.  The inferred indices are $\beta = -0.87 \pm 0.02$ ($\chi^2_r = 3.6/3 = 1.2$) and $\beta = -0.85 \pm 0.02$ ($\chi^2_r = 2.6/3 = 0.9$) for Models A and B, respectively.  Uncertainties in the unabsorbed continuum are evaluated filter by filter; centering the frequencies at each filter frequency $\nu^\prime$ allows us to avoid covariances with the uncertainty in the SED slope.  Namely, for
\begin{equation}
\ln{F_\nu} = \beta ~\ln{x} + \ln{F_{\nu,0}},
\end{equation}
where $x = \nu/\nu^\prime$, $\Delta F_\nu = \Delta F_{\nu,0}$ when $x=1$.

\subsection{Deriving the Extinction Curve}
\label{subsect:extcurve}

For Models A and B, the 7 optical and UV data points lie below the SED fit, indicating extinction by dust present within the host galaxy of GRB 050525A.  The extinction in a given waveband is related to the optical depth, $\tau_\lambda$, by $A_\lambda = 2.5 \tau_\lambda / \ln{10}$; we calculate $A_\lambda$ for our 7 optical/UV wavebands and present them in Table \ref{tab:ext}.  The derived dust extinction curves are shown in Figs. \ref{fig:model_a} and \ref{fig:model_b} for Models A and B, respectively.  In general, the extinction curve is a steeply-increasing function of the photon frequency in the rest frame of the host galaxy, $\nu_0$, and can be fitted with a power law: the resulting indices are $\delta = 1.01 \pm 0.31$ ($\chi^2_r = 5.6/5 = 1.1$) and $\delta = 1.23 \pm 0.40$ ($\chi^2_r = 5.5/5 = 1.1$) for Models A and B, respectively.  These fits are noticeably steeper than the $\tau_\lambda \propto \nu^{0.7}_0$ attentuation curve derived by Johnson et al. (2007).  Broad features, such as the 2175 \AA\ bump seen in our Galaxy and the LMC, do not appear to be present (but see \S\ref{subsect:dustgas}).  Starling et al. (2007) derived similar results for their sample of 10 GRB hosts using known extinction law templates.

We note that if we only use the 3.6 and 4.5 $\mu$m flux measurements when fitting a power law function to the SED, then our results are consistent with zero extinction in the $V$ and $R$ bands.  We tabulate the extinction values resulting from such a scenario as ``Model C'' in Table \ref{tab:ext}.  In the other extreme, we explore a ``Model D'', where we instead use only the 8.0 and 24.0$\mu$m (and ignore the other IR) data points, which serves as a conservative upper limit on the amount of extinction present.

We caution that the Galactic extinction towards GRB 050525A is comparable to, and in some cases larger than, our derived intrinsic extinction.  Given the possible variations in the Galactic extinction curve, it is possible that additional sources of errors are present which we have not considered.  For example, Schlegel, Finkbeiner \& Davis (1998) report a Galactic $V$-band extinction of $A^{(g)}_V = 0.316$, but with an error of $\Delta A^{(g)}_V \sim 0.1$ or $\sim 30\%$.  Other sources of uncertainty include the Galactic UV bump (Savage et al. 1985) and the assumed $R_V$ value for the MW (Cardelli, Clayton \& Mathis 1989; and references therein).

\subsection{Obtaining Dust-to-Gas Ratios for Various Extinction Laws}
\label{subsect:dustgas}

We now wish to study the consistency of our results with the MW, SMC and LMC extinction laws.  These extinction laws were previously presented in Weingartner \& Draine (2001) and Li \& Draine (2001), but have since been updated in Draine (2003a, b, c).  We first convolve the model extinction laws with the relevant filter responses, which for simplicity are assumed to be Heaviside (i.e., ``top hat'') functions.  The filter widths are taken to be 128, 75, 98, 88, 70, 51 and 76 nm for the $R$, $V$, $B$, $U$, $UVW1$ (260 nm), $UVM2$ (220 nm) and $UVW2$ (198 nm) bands, respectively.  We then fit for the optimal value of the host galaxy hydrogen column density\footnote{Specifically, the hydrogen-equivalent column density intrinsic to the host galaxy, along the line of sight to the GRB.  This could be different if GRBs occur in special regions of galaxies.}, $N_{\rm H}$, that allows the model law ($A_\lambda/N_{\rm H}$) to be matched to the observed curve ($A_\lambda$).  We next use our power law fits to the extinction curves (see \S\ref{subsect:extcurve}) to derive the value of the visual extinction in the $V$ band, $A_V$, {\it in the rest frame of the host galaxy}: $A_V = 0.22 \pm 0.06$ (Model A), $0.15 \pm 0.06$ (Model B), $0.02^{+0.03}_{-0.02}$ (Model C) and $0.41 \pm 0.14$ (Model D).  The resulting $A_V/N_{\rm H}$ values for Models A and B are reported in Table \ref{tab:avnh}.

We first see that the MW extinction laws, regardless of the $R_V$ value and the assumption of either Model A or B, are somewhat disfavored even allowing for different dust-to-gas ratios; less dust is generally required than expected.  This conclusion is also true for Models C and D.  We also tabulate the ratio of our derived $A_V/N_{\rm H}$ value to the theoretically expected one for the various extinction laws, a quantity we denote by $f_d$.  We see that the SMC and LMC laws provide good fits to our extinction curve.  Based on the derived $f_d$ values, about 10 to 40\% more dust is present if we use a SMC fit.  For the LMC fit, Model A requires 10\% more dust, while Model B needs 20\% less dust.  The similar $\chi^2_r$ values of the SMC and LMC fits do not allow us to rule out the possibility that an LMC-like 2175 \AA\ bump may be present in our extinction curve.  We note that Models C and D favor the SMC law over the LMC one.

In general, the dust-to-gas ratio found in the host galaxy of GRB 050525A resembles that found in the LMC, and is intermediate between the ratios for the SMC and our Galaxy.  It is also similar to that found in the circum-nuclear region of AGNs (Maiolino et al. 2001).  The relationship between $N_{\rm H}$ and $A_V$ has been studied in GRB hosts by Galama \& Wijers (2001) and Stratta et al. (2004). Their derived dust-to-gas ratios are approximately ten times smaller than in the MW, which is roughly consistent with our findings.  It is comforting that our derived values for the extinction curve and $A_V/N_{\rm H}$ are somewhat similar for Models A and B.

\section{Discussion}
\label{sect:discussion}

\subsection{Comparison to Other Studies}

Previous studies of the dust extinction laws in the circum-burst
environment of GRBs used known extinction templates (MW, LMC and SMC).
Assuming Galactic dust conditions, Galama \& Wijers (2001) found
$N_{\rm H} \sim 10^{21}$ to $10^{23}$ cm$^{-2}$ and low extinctions
($A_V \sim 0.7$) for their sample of 8 GRBs.  For 13 GRB afterglows,
Stratta et al. (2004) examined several extinction laws and inferred
similar column density values that are generally larger than expected
for the given $A_V$. In most cases steep, SMC-like
extinction curves were preferred (Starling et al. 2007), although in some cases flat, gray curves
have been claimed (e.g., Stratta et al. 2005).  Savaglio, Fall \& Fiore
(2003) infer $A_V \approx 1$ mag, in contrast to $A_V \lesssim 0.1$
mag found in damped Ly$\alpha$ (DLA) systems along QSO sight lines,
supporting the idea that optically-selected QSOs mainly probe regions of high-$z$ galaxies with less dust and smaller gas column densities, while GRBs probe the
denser, star-forming, ISM regions, and thus the two approaches are
complementary.  (See also Savaglio \& Fall 2004.)

The {\it Spitzer} IR observations of GRB 050525A allow us to
determine the extinction curve in its host galaxy while making less assumptions than in previous studies.  Recently, Li, Li \& Wei (2007) applied a similar but rather risky methodology, where they attempted to determine the spectral slope of the unabsorbed continuum by employing extrapolation from the X-ray to the optical range of wavelengths.  Lacking the IR information means that their results are subjected to large uncertainties.  In addition, the degeneracy of
the X-ray spectral slope with the amount of soft X-ray photo-absorption present
makes the computation of $A_V/N_{\rm H}$ extremely challenging.  In our
treatment, the X-ray spectral information is used only as a
consistency check.  Li, Li \& Wei (2007) claimed that the GRB hosts they study have gray extinction curves, but this is not surprising considering their basic assumption regarding the slopes of their SEDs.

\subsection{Implications for Host Galaxy Studies}

A detailed study of the properties of the host galaxy of GRB 050525A is beyond the scope of our paper, but we briefly discuss some implications of our results to such a task.  The origin of the 2175 \AA\ bump is a subject of controversy --- Draine \& Li (2007) suggest that it is due to the presence of polycyclic aromatic hydrocarbons (PAHs) in the dust population.  However, these molecules have been found to be deficient in low-metallicity, high radiation-density, blue dwarf galaxies, the favored hosts of GRBs (e.g., Hogg et al. 2005; Engelbracht et al. 2008).  If our derived extinction curve can be more closely sampled in frequency, it will allow us to make a more definitive statement regarding the absence/presence of the 2175 \AA\ bump, which may shed light on the PAH hypothesis.  In their study of dwarf galaxies, Engelbracht et al. (2008) inferred dust-to-gas ratios that differ by 3 orders of magnitude, preventing a meaningful comparison with our derived $A_V/N_{\rm H}$ values.

D06 analyzed the optical spectrum of GRB 050525A at the time of the SN
bump and concluded that SN 2005nc is akin to SN 1998bw, an energetic
Type Ic SN (Patat et al. 2001), but dimmed by $\sim 0.3$ mag.  If we
speculate that SN 2005nc is a Type Ic event as well, it is interesting
to ask if GRB 050525A/SN 2005nc erupted in the brightest region of a
spiral or irregular galaxy, where conditions are appropriate for
star formation, as suggested by Kelly, Kirshner \& Pahre (2007), and if the inferred dust-to-gas content is consistent with such an environment.  Such
a determination is relevant to the hypothesis that metal-poor and
metal-rich Wolf-Rayet stars, the proposed progenitors of Type Ic
supernovae, result in long GRBs and less energetic core collapse
explosions, respectively (Yoon \& Langer 2005; Woosley \& Heger 2006).

\subsection{Is the Derived Extinction Curve Pristine?}

The dust extinction curve at late times (i.e., during the afterglow phase) may be indicative of the residual extinction present after the onset of the burst radiation.  The effects of dust sublimation can be dramatic if the absorbing region is small and compact (e.g., in a molecular cloud), while they are negligible for typical ISM densities (Waxman \& Draine 2000).  A direct way to quantify the modification of the dust grain population by the burst radiation is to measure the optical opacities at very early times, when these changes are occuring.  However, these times are generally too short for the effects to be observable (Perna et al. 2003).

If the absorbing region is small and compact, then variations in the optical opacities should be accompanied by changes in the effective column density due to the ionization of the gas.  These effects occur on a reasonably long timescale and are detectable (e.g., Lazzati \& Perna 2002, 2003; Campana et al. 2007).  To within the measurement uncertainties, B06 detected no changes in $N_{\rm H}$ from $t = 250$ to 25,000 s (see their Table 6).  To accomodate a situation in which dust is destroyed without an accompanying, significant ionization of the gas requires specially fine-tuned conditions (Perna \& Lazzati 2002).  We therefore conclude that our measured extinction curve is pristine to the host galaxy of GRB 050525A and was not modified by the prompt emission of the GRB.

\subsection{Future Work \& Opportunities}

Our method of deriving the dust extinction curve of a GRB host galaxy is comparatively robust, since we have cut down on the number of assumptions made in previous studies.  Nevertheless, we regard our work as a ``proof of concept'' study, which allows extinction curves to be derived independently of the known templates.  Future work may improve on ours by not only obtaining more IR and optical/UV observations simultaneously (so as to avoid the fitting procedure described in \S\ref{sect:lightcurves}) and earlier in time, but also sampling the SED at more frequencies in between the IR and the X-ray.  Such observations should be complemented by studies of the host galaxy emission.  For example, the {\it Gamma-Ray Burst Optical Near-Infrared Detector (GROND)}, a 2.2 m European telescope, is capable of observing GRB afterglows simultaneously in 7 wavebands from 400 nm to 2.2 $\mu$m, and is ideally suited for the type of study we have presented in this paper, provided contemporaneous X-ray measurements can be coordinated.

The presence of the 2175 \AA\ bump can be inferred regardless of the normalization of the unabsorbed continuum or its power-law slope.  However, these two pieces of information are required to derive the shape of the extinction curve and the $A_V/N_{\rm H}$ values, and therein lies the power of the IR data.  Besides {\it GROND}, near-infrared (NIR), ground-based observatories will play a key role in such endeavors.

\acknowledgments
\scriptsize
K.H. gratefully acknowledges support from the
Institute for Advanced Study and benefits from diverse interactions
within the astrophysics group.  He thanks Brian Schmidt, Alexia Schulz, Mike Fall, Nadia Zakamska, Alex Blustin and Ralph Wijers for helpful conversations.  This work
was partially supported by NSF grant NNG05GH55G (R.P. and D.L.).  We are indebted to the anonymous referee for his/her meticulous reading of the manuscript and the subsequent, useful suggestions that followed, which tremendously improved the quality of the paper.
\normalsize



\begin{table}
\begin{center}
\caption{Optical/ultraviolet/X-ray Fits \& Infrared Observations of GRB 050525A}
\label{tab:spitzer}
\begin{tabular}{lcccc}
\tableline\tableline
\multicolumn{1}{c}{Band} & \multicolumn{1}{c}{Instrument} & \multicolumn{1}{c}{Date} & \multicolumn{1}{c}{Flux}\\
\multicolumn{1}{c}{} & \multicolumn{1}{c}{} & \multicolumn{1}{c}{} & \multicolumn{1}{c}{$\mu$Jy}\\
\tableline
2 keV & {\it XRT/Swift} & $\diamond$ & $\left(7.22^{+0.96}_{-0.82}\right)_{-3}$ \\
198 nm & {\it UVOT/Swift} & $\diamond$ & $\left(3.40^{+0.27}_{-0.41}\right)_{-1}$ \\
220 nm & {\it UVOT/Swift} & $\diamond$ & $\left(6.60^{+0.64}_{-0.66}\right)_{-1}$ \\ 
260 nm & {\it UVOT/Swift} & $\diamond$ & $\left(6.60^{+0.81}_{-0.79}\right)_{-1}$ \\
U & {\it UVOT/Swift} & $\diamond$ & $1.15^{+0.12}_{-0.14}$ \\
B & {\it UVOT/Swift} & $\diamond$ & $1.20^{+0.09}_{-0.15}$ \\
V & {\it UVOT/Swift} & $\diamond$ & $1.90^{+0.14}_{-0.20}$ \\
R & Various & $\diamond$ &  $2.40^{+0.15}_{-0.21}$ \\
3.6 $\mu$m & {\it IRAC/Spitzer} & May 27.264 &  $8.4 \pm 3.3$ \\
4.5 $\mu$m & {\it IRAC/Spitzer} & May 27.264 &  $10.7 \pm 4.2$ \\
5.8 $\mu$m & {\it IRAC/Spitzer} & May 27.264 &  $17.9^\dagger$\\
8.0 $\mu$m & {\it IRAC/Spitzer} & May 27.264 & $32.2 \pm 9.3$ \\
24.0 $\mu$m & {\it MIPS/Spitzer} & May 27.722 & $104.8 \pm 50.0^\ddagger$ \\
\tableline
\end{tabular}
\end{center}
\scriptsize
Note: $A_b$ is shorthand for $A \times 10^b$.\\
$\diamond$: Light curve fit to May 27.264.\\
$\dagger$: Indicates an upper limit.\\
$\ddagger$: Scaled to $t=t_{\rm IR}$.\\
\normalsize
\end{table}

\begin{table}
\begin{center}
\caption{Derived dust extinction in magnitudes}
\label{tab:ext}
\begin{tabular}{lcccc}
\tableline\tableline
\multicolumn{1}{c}{Band} & \multicolumn{1}{c}{$A_\lambda$ (Model A)} & \multicolumn{1}{c}{$A_\lambda$ (Model B)} & \multicolumn{1}{c}{$A_\lambda$ (Model C)} & \multicolumn{1}{c}{$A_\lambda$ (Model D)}\\
\tableline
$R$ (0.71 $\mu$m) & $0.30^{+0.18}_{-0.17}$ & $0.21^{+0.18}_{-0.17}$ & $\dagger$ & $0.53^{+0.22}_{-0.21}$ \\
$V$ (0.55 $\mu$m) & $0.32^{+0.19}_{-0.17}$ & $0.23^{+0.19}_{-0.17}$ & $\dagger$ & $0.54^{+0.22}_{-0.21}$ \\
$B$ (0.44 $\mu$m) & $0.61^{+0.20}_{-0.17}$ & $0.52^{+0.20}_{-0.17}$ & $0.28^{+0.26}_{-0.24}$ & $0.82^{+0.23}_{-0.20}$ \\
$U$ (0.36 $\mu$m) & $0.46^{+0.19}_{-0.18}$ & $0.38^{+0.19}_{-0.18}$ & $0.15^{+0.26}_{-0.15}$ & $0.67^{+0.22}_{-0.21}$ \\
$UVW1$ (260 nm) & $0.76 \pm 0.19$ & $0.68 \pm 0.19$ & $0.46 \pm 0.25$ & $0.96 \pm 0.22$ \\
$UVM2$ (220 nm) & $0.60 \pm 0.17$ & $0.53 \pm 0.17$ & $0.31 \pm 0.23$ & $0.80 \pm 0.20$ \\
$UVW2$ (198 nm) & $1.22^{+0.19}_{-0.16}$ & $1.15^{+0.19}_{-0.16}$ & $0.94^{+0.24}_{-0.22}$ & $1.41^{+0.21}_{-0.19}$ \\
\tableline
\end{tabular}
\end{center}
\scriptsize
$\dagger$: Consistent with zero extinction.
\normalsize
\end{table}

\begin{table}
\begin{center}
\caption{Derived visual extinction per column density}
\label{tab:avnh}
\begin{tabular}{lcccccc}
\tableline\tableline
\multicolumn{1}{c}{Model} & \multicolumn{1}{c}{$A_V/N_{\rm H}$ (Model A)} & \multicolumn{1}{c}{$\chi^2_r$ (Model A)} & \multicolumn{1}{c}{$f_d$} & \multicolumn{1}{c}{$A_V/N_{\rm H}$ (Model B)} & \multicolumn{1}{c}{$\chi^2_r$ (Model B)} & \multicolumn{1}{c}{$f_d$}\\
\multicolumn{1}{c}{} & \multicolumn{1}{c}{$(10^{-22} \mbox{ cm}^{2})$} & \multicolumn{1}{c}{$\left(N_{\rm dof} = 6\right)$} & \multicolumn{1}{c}{} & \multicolumn{1}{c}{$(10^{-22} \mbox{ cm}^{2})$} & \multicolumn{1}{c}{$\left(N_{\rm dof} = 6\right)$} & \multicolumn{1}{c}{} \\
\tableline
MW $\left(R_V = 3.1\right)$ & $(4.44 \pm 1.41)_{-22}$ & 1.4 & 0.8 & $(3.36 \pm 1.37)_{-22}$ & 1.5 & 0.6 \\
MW $\left(R_V = 4.0\right)$ & $(4.10 \pm 1.34)_{-22}$ & 1.9 & 0.7 & $(3.12 \pm 1.30)_{-22}$ & 2.1 & 0.5 \\
MW $\left(R_V = 5.5\right)$ & $(3.81 \pm 1.30)_{-22}$ & 2.5 & 0.5 & $(2.92 \pm 1.26)_{-22}$ & 2.7 & 0.4 \\
SMC & $(8.55 \pm 2.68)_{-23}$ & 1.1 & 1.4 & $(6.37 \pm 2.52)_{-23}$ & 1.0 & 1.1 \\
LMC & $(1.30 \pm 0.40)_{-22}$ & 0.9 & 1.1 & $(9.76 \pm 3.88)_{-23}$ & 1.0 & 0.8 \\
\tableline
\end{tabular}
\end{center}
\scriptsize
Note: $A_b$ is shorthand for $A \times 10^b$.\\
\normalsize
\end{table}

\end{document}